\documentclass[a4paper]{article}
\usepackage[T1]{fontenc}
\usepackage[utf8]{inputenc}
\usepackage{amsthm,amsmath,amsfonts,amssymb}
\usepackage{pdfsync}
\usepackage{enumerate}
\usepackage{graphicx}
\usepackage{epsfig} 
\usepackage{calc}
\usepackage{textcomp}
\usepackage{tabularx}
\usepackage{txfonts}
\usepackage{hyperref}
\usepackage{physics}
\usepackage{color}
\usepackage{authblk}
\usepackage[authordate]{biblatex-chicago}
\title{Robustness and the Event Horizon Telescope: the case of the first image of M87*}

\author[1,3]{Juliusz Doboszewski}
\author[2,3]{Jamee Elder}
\affil[1]{Lichtenberg Group for History and Philosophy of Physics - Institute of Philosophy, University of Bonn, 53113 Bonn, Germany}
\affil[2]{Department of Philosophy - Tufts University, Medford, MA 02155, USA}
\affil[3]{Black Hole Initiative - Harvard University, Cambridge, MA 02138, USA}

\begin{document}

\maketitle

\begin{abstract}
We examine the justification for taking the Event Horizon Telescope's famous 2019 image to be a reliable representation of the region surrounding a black hole. We argue that it takes the form of a robustness argument, with the resulting image being robust across variation in a range of data-analysis pipelines. We clarify the sense of ``robustness'' operating here and show how it can account for the reliability of astrophysical inferences, even in cases---like the EHT---where these inferences are based on experiments that are (for all practical purposes) unique. This has consequences far beyond the 2019 image.
\end{abstract}

\tableofcontents

\section{M87*: on the need for a philosophy of the 2019 EHT image}

In April 2017 eight radio telescopes in six geographical locations were used by the Event Horizon Telescope collaboration as elements of a virtual telescope with an aperture of approximately the diameter of the Earth. This allowed the collaboration to achieve an angular resolution of 20$\mu$as, making the EHT array one of the highest resolution astronomical instruments to date, as well as the first (and only) instrument capable of imaging the shadow of a supermassive black hole.\footnote{This is comparable to the resolving power needed to observe an orange on the surface of the moon, assuming it emitted radiation at 230 GHz. An orange at this distance appears to be a similar size to the closest supermassive black hole, Sagittarius A* (SgrA*), at a distance of about 8kpc. Despite differences in mass between M87* and SgrA*, their respective distances from Earth imply that they are of comparable angular size (approximately $51\mu$as for SgrA* and $42\mu$as for M87*).} This array was used to perform a precision measurement of the radio core of the Messier 87 Galaxy (``M87''). In April 2019 the resulting image of that source was widely publicized. The image was found to be consistent with a rotating black hole described by the Kerr solution of the general theory of relativity, providing a new test of the theory as well as one of the best pieces of evidence for the existence of black holes to date.

The first image of the core of M87\footnote{For simplicity, we will follow the convention of denoting the core of M87 as M87*.} is a natural and exciting target for philosophical analysis, but interest in the EHT also extends well beyond this particular image. Already the 2017 data have been used for other imaging projects, including the jet of the blazar 3C 279 (\cite{2020EHT3C279}), linear polarization of M87* (\cite{2021EHT_M87_paper7}, \cite{2021EHT_M87_paper8}), and the jet in Centaurus A (\cite{2021EHTcenAjet}). Moreover, the 2017 observations included measurements of Sagittarius A* (SgrA*), the supermassive black hole candidate at the center of the Milky Way.\footnote{The SgrA* results have recently been released; see \textcite{EHT2022_SgrA_paper1}. There are important differences between M87* and SgrA*, in particular concerning the timescales of their variability, and so we defer detailed discussion of the new SgrA* image to future work. However, the overall imaging procedures are similar, and so our analysis of the role of robustness in justifying the 2019 image of M87* carries over to the 2022 image of SgrA*, at least in broad strokes.} Further observational runs in 2018, 2019, 2021, and 2022 also covered M87* and SgrA*, among other targets. During this time, three new stations have been added to the array and a plan for a custom, less heterogeneous array (the next generation, so ngEHT) has begun to emerge; see \textcite{ngeht-key-science-goals} and \textcite{ngeht-hpc-2023} for the broader philosophical perspective as well as the way in which humanities scholar will be embedded in that collaboration. In the far future, another significant jump in resolution could be made with the addition of space components in Earth's orbit or even further in outer space.\footnote{Space VLBI has been successfully demonstrated by the Haruka telescope in 1997-2005, and more recently with the Spektr-R satellite (2011-2019) within the RadioAstron program; \textcite{gurvits2020space} is a useful summary of these missions.}

Since these imaging projects (past and future) employ very similar methods to the imaging of M87*, we expect our analysis to straightforwardly extend to those cases. We also expect that our analysis of the role of robustness in justifying the EHT results for M87* will be relevant to understanding experiments that share relevant epistemic features with the EHT.\footnote{The choice to describe the EHT observations as an ``experiment'' is potentially controversial, since others may prefer to distinguish between ordinary experiments, where phenomena are produced by interventions of the experimenters on the target system (i.e., the system of interest), and observations, where the scientist collects (causally downstream) information about a target system that they do not (or cannot) intervene on. Such a distinction has clear connections with the challenges of justifying the EHT results---challenges that robustness analyses are intended to mitigate. It is beyond the scope of this paper to evaluate the usefulness of distinguishing between experiments and observations in this manner, but we choose to use `experiment' terminology for two key reasons: (1) in order to emphasize the similarities in epistemic situation across large experiments like the EHT, LIGO-Virgo, and the LHC; and (2) in order to resist the common, if implicit, presumption of the epistemic superiority of experiments over `mere' observations.} This includes large experiments in other fields (section \ref{epistemic_situation_large_experiments}) as well as other astronomical experiments (section \ref{interferometry_VLBI}). We situate our analysis relative to philosophical work on robustness in other experiments (e.g., at the Large Hadron Collider and in cosmology) in section \ref{conclusions}.

When considered as an experiment, the EHT observations are rather unusual. Varying the experimental setup or the population of targets to get varied evidence about the target is not a viable option for the EHT: there is no experimental control over such targets, there is no alternative similar array available, and presently the number of sensible targets is small.\footnote{\label{fn:calibration_sources}In 2017, the primary targets were Sgr A* and M87*, and the secondary targets were and active galactic nuclei OJ 287 (a candidate for a supermassive black hole binary system), Centaurus A, NGC 1052, and blazar 3C 279. Some of these served as calibration sources for others.} Due to its unprecedented resolution, the data gathered by the EHT provide a unique line of evidence about the near horizon-scale structure of a supermassive black hole.\footnote{However, it is \textit{not} a unique line of evidence for other purposes, such as measurement of the mass of the central object \autocite[8.2]{2019EHT_M87_paper6}.} 
 
The trustworthiness of the instrument and the reliability of the data analysis methods are crucial for trust in the outcome, and understanding the structure of justification behind it is of high philosophical importance. But what are sources of confidence in situations where there is no variation in the initial conditions, in population, or in tools used? What reasons might we have for considering the 2019 EHT images to be reliable representations of the supermassive black hole candidate in the center of the M87 galaxy? Do these reasons take the form of a robustness argument? If so, what does it involve, and in what ways is it similar to other robustness arguments? More generally, what can be done in order to ensure the reliability of results obtained from experiments such as the EHT observations?

In order to address these questions, we will discuss a few types of epistemological concerns. Those come from very different sources: some are specific to large experiments, others to radio astronomy, yet others to the EHT itself, and finally some to astrophysics more generally. First we consider the epistemic peculiarities of what we call ``large experiments''---physically big, expensive, and unique experiments at the cutting edge of empirical investigation. We will then (in section \ref{interferometry_VLBI}) discuss some of the specific methodological and epistemological issues arising in radio-astronomy in general, and the EHT in particular.
It turns out that the EHT collaboration explicitly makes use of a robustness argument in establishing the validity of their conclusions. Accordingly, in section \ref{robustness} we survey available notions of robustness and their applicability to astrophysical situations.
Section \ref{EHT_2019} follows up on that with a reconstruction of the EHT Collaboration's main lines of reasoning, and an analysis of how variation in the EHT methods---especially during imaging---allows them to claim that their result is ``robust'' and hence reliable. In section \ref{conclusions} we use our analysis to clarify some confusion concerning use of robustness arguments in black hole imaging.

\section{Epistemic peculiarities of large experiments}\label{epistemic_situation_large_experiments}
The EHT is a large, cutting-edge experiment, with similarities to other large-scale experiments such as the LIGO-Virgo gravitational-wave detectors, and high-energy physics experiments conducted at the Large Hadron Collider (LHC).\footnote{Other examples include the ocean observatory NEPTUNE with its 840 km underwater fiber optic cable loop, the EarthScope network of seismometers, the IceCube neutrino detector at the South Pole; the list goes on.} Such experiments have some epistemic peculiarities compared to the typical small-scale ``tabletop'' experiments undertaken by small groups of scientists working in ordinary laboratory settings.

In using the term ``large'' to describe an experiment, we have in mind a cluster of related features. An experimental or observational setup may be physically \textit{big}, meaning that the instrumentation itself is large (as in the cases of the LHC and LIGO), that components of the instruments are spread out over large distances (as in the cases of LIGO and the EHT). Such a physical setup may also be \textit{expensive} in that building and/or operating it requires a high budget; additionally, a large group of people may be required to operate it and run the observations.

These constraints tend to make such experiments \textit{cutting edge} in that the setup is used for making observations in previously unexplored regimes---be it previously inaccessible energy scales in the case of the LHC, dynamical processes in the strong field regime observed with LIGO-Virgo or the resolution achieved by the Event Horizon Telescope. This can be seen as a consequence of their source and resource expenditure: the cost of the construction and operation is justified by the new insights expected to be obtained through the experiment. These features are generally realized by a particular experiment as a matter of degree (for example, one experimental setup can be larger than another). 

The experiments we have in mind also tend to be \textit{unique} in the sense that a second such experiment is not available.\footnote{This kind of uniqueness, is, of course, something highly contingent. For example, the detection of the Higgs boson at the LHC involved two independent detectors, Atlas and CMS; these are not two fully independent experiments, but they do provide some form of independence. Hypothetically, a cold war-type scientific race could lead to the development of competing large experiments. Future technological developments can also render previously cutting-edge experimental setups commonplace.} Reasons for that vary. Constructing another instrument may come with a high price tag.\footnote{For LIGO-Virgo, this amounts to 1.1 billion USD over the four decades leading up to the 2015 detection, and in the LHC case this amounts to 1.1 billion CHF for operational costs in 2009-2012 alone; the next generation LHC is estimated to cost between 9 and 21 billion EUR, depending on the design. Although in absolute terms (or compared to many defense budgets) these numbers are not particularly impressive, they are extremely high in the context of basic science funding. One could also reasonably expect that a firm commitment by either CERN or Chinese IHEP to a next-generation LHC detector might effectively nullify the willingness of the other organization to build a second detector, as it likely would be seen as redundant.} The size of the instrument or need for a very special location may also be reasons why an experiment is unique\footnote{For instance, locations for the Atacama Large Millimeter Array (ALMA) or the South Pole Telescope have been chosen partially on the basis of high elevation and low humidity. For space-based instruments an example is Lagrange points: these are points where stable orbits are much easier to achieve than in other locations. For gravitational wave astrophysics, flat, remote sites with minimal background noise sources are needed. For detailed history of the site selection for the LIGO interferometers, see \textcite{Nichols_Diss,Nichols2017,Nichols2021}}. Large experiments may also involve a large part of the relevant scientific sub-community, limiting available human resources and making a second such experiment practically impossible.

This cluster of properties makes the epistemic situation of large experiments quite interesting. Large experiments have clear advantages, but they also face special challenges. 

On the one hand, they provide empirical data that cannot be collected using smaller setups. Indeed, large and expensive projects are undertaken because of the potential importance of the data they can provide. The outcomes of such experiments also represent the consensus of a large group of scientists. Ideally, this collaborative process limits the impact of individual researchers' biases on experimental outcomes.\footnote{This is contingent upon various features of the way that the collaboration is set up. For example, \textcite{Zollman2007,Zollman2010} shows that under certain circumstances increased communication between scientists can lead to erroneous experimental results being accepted by the group (see also \textcite{rosenstock_bruner_o’connor2017}). In another context, \textcite{Marcoci-Nguyen2020} show that unanimity preservation can indirectly come into conflict with expertise (see also \textcite{Bright-Dang-Heesen2018}).}

On the other hand, replicability of the results of such experiments is limited in several ways. For an experiment that is both unique and cutting edge, alternative ways of investigating the  physical regime are simply not available. Furthermore, issues such as cost and lack of availability of trained technicians might ensure that a unique experiment remains so. Finally, some large experiments rely on the challenging logistical task of adapting and coordinating existing instruments for a new purpose. For instance, the EHT array comprises several existing radio-telescopes. Providing these telescopes with the required additional equipment (such as hydrogen masers used to time stamp the recorded data) and obtaining coordinated observational time on all of them simultaneously presents a major logistical challenge. It would be far from trivial for another collaboration to replicate such a feat, even using the same components. Subsequent observations by the existing collaboration merely provides us with more data from the same instrument, collected by a very similar group of people in roughly the same way; something quite different from a confirmation coming from another group using a different instrument. It may seem (to paraphrase \textcite[153]{Cartwright1991}) that this would amount to doing the same thing in the same way twice.\footnote{See section \ref{varied_models} for a discussion of Cartwright's views and \ref{conclusions} for our discussion of these issues with respect to the EHT specifically.} In effect, this would fall short of a certain ideal of how science should be operating---one where results can be \textit{reproduced} by independent groups conducting independent experiments.

Throughout this paper we will argue that a strategy based on robustness, i.e. relying on a plurality of data analysis methods, bolsters confidence in the conclusions reached by the EHT by providing convergent and discriminant validation (in the sense of \textcite{Staley2004robustness}). Varying data analysis pipelines can also be a more general viable strategy for ensuring the reliability of large scale experiments, where (for reasons including the physical scale, cost, etc. outlined above) replicating the experiment or otherwise gaining independent access to the target phenomena may be unrealistic.
Another strategy would involve choosing a single analysis method and working to improve confidence in that method alone. While such a strategy is important, it will not always be sufficient, especially given that lack of independent access to the target restricts available benchmarks. In other words, for large scale experiments there may be limits on the extent to which one can demonstrate that a given method is reliable in the relevant domain. 
Relying on a single method in such a situation becomes risky, as overconfidence in that method might lead to accepting a false conclusion (see the BICEP2 case discussed in section \ref{astrophysics}). If one cannot reduce the uncertainty in a single method any further, then what can be done is to show insensitivity of the overall conclusion to those uncertainties. This is precisely what robustness analysis does.
The situation changes once a domain under investigation is no longer novel, but has been to some extent explored: in that situation, results of a single analysis method can be bench-marked against previous outcomes, and the role of the robustness-based strategy diminishes. Thus, robustness strategies seem to be most important on scientific frontiers, when new experiments (often of large-scale compared to previous experiments in that field) probe new phenomena and must manage new uncertainties without established benchmarks.  

With an array spanning the Earth, '(t)he EHT achieves an extraordinary resolution of 13 $\mu$as, making it the highest resolution imaging instrument in the history of astronomy' \autocite[3]{broderick2020themis}.\footnote{The higher resolution stated here (13 $\mu$as as opposed to the 20 $\mu$as stated in the main EHT papers) reflects the fact that determining the resolution is a complex business. As noted below, eht-imaging and SMILI are methods that produce higher-resolution images than the traditional CLEAN method in this setting. The blurring of these images to a conservative resolution of 20 $\mu$as to match the CLEAN image is a particular epistemic choice in response to questions about which features of the final images are trustworthy. We discuss these issues in connection with Kent Staley's notion of a weakening strategy in section \ref{EHT_2019}.} 
These features make it both unique and cutting edge---decidedly a large experiment in our sense. As systematic variation of the experimental setup or target systems to get varied evidence is simply not a viable option here, the data gathered by the EHT provide a unique line of evidence about the horizon-scale structure of a supermassive black hole. This further exacerbates the challenging epistemic situation in astrophysics (briefly reviewed in section \ref{astrophysics}) as well as significant epistemological worries about the particular experimental methods being applied---in this case, those of radio interferometry, specifically Very Long Baseline Interferometry. 

\section{Basics of interferometry and VLBI in radio astronomy}\label{interferometry_VLBI}

Almost every astronomical source can be thought of as a thermal emitter, projecting radio waves at various frequencies. Interferometry is one of the basic ways of measuring the brightness of a source.\footnote{Our discussion here follows the standard reference \textcite{Thompsonetal2017} (especially chapters 3, 5 and 9). Readers interested in more details could also consult \textcite{condon2016essential} for a more concise introduction to the basics of radio astronomy (which, however, does not go into the details of VLBI or imaging algorithms), as well as Science and Technology sections of the Event Horizon Telescope website for an illustration of the methods EHT uses.}

A single interferometer consists of two apertures, separated by some distance $B$, the baseline. Apertures receive light beams from the source; these are then combined in the receiver, producing a fringe pattern. The response of an interferometer is a function of the source brightness, fringe separation, and orientation of the device. If the angular width of the source is comparable to the distance between the light intensity maxima received in the apertures, the resulting image arises from a superposition of snapshots covering the source. Maxima and minima of the interference fringes from distinct points do not coincide, attenuating the fringe amplitude. The fringe visibility V is then defined as (intensity of maxima - intensity of minima) divided by (intensity of maxima + intensity of minima. $V = 1$ if intensity at the minima is zero, \textit{i.e.}, when the width of the source is comparable with the fringe width. Wider bandwidths and longer baselines result in narrower fringe envelopes. A source is resolved by the interferometer if $V < 1$. A single interferometer measures a one-dimensional intensity profile; to obtain an image of a two-dimensional source with intensity profile $I(l, m)$, a radio interferometer array is needed. $(l, m)$ are Cartesian coordinates on the sky which are chosen in such a way that $l$ is measured parallel to the aperture spacing and $m$ is normal to it.

The angular resolution (in radians) of a resolved source (so the resolution of a final image) is $\theta = \frac{\lambda}{B}$, where $\lambda$ is the wavelength of a radio wave and $B$ is the length of a baseline. A commonly used coordinate system for describing baselines uses, again, the Cartesian coordinates $(u,v,w)$, oriented in such a way that the plane containing the antennas is located at $w=0$. This frame is used to measure the baseline components, not the locations of the antenna. Obtaining more precise images, or any images of very distant sources, requires longer baselines $B$ or shorter wavelengths $\lambda$; of particular interest for us are millimeter arrays (1.3 mm for the EHT). For distant sources such observations require the use of the very-long-baseline interferometry (VLBI), consisting of arrays of antennas that are separated by hundreds of kilometers (or more). The EHT reaches an angular resolution of 20$\mu$as by incorporating baselines approximating the diameter of the Earth (in 2017 the longest baseline was 10700 km long; see figure 1 in \cite{2019EHT_M87_paper1} for a schematic depiction of the array).

An interferometer measures fringe visibilities: two-point correlation functions of the electric field emitted by the source. According to the Cittert–Zernike theorem, the visibility function $V(u, v)$ (in $(u,v)$ plane coordinates) is the Fourier transform of the source intensity distribution $I(l, m)$. This means that when pairs of antennas in an array measure fringe visibilities, they are sampling the Fourier components of the source emission.\footnote{See chapter 15 of \textcite{Thompsonetal2017} for an extended introduction.}

Reconstruction of an image requires calculating the intensity distribution of a source on the sky from the measured visibilities. This is an example of an inverse problem, which in general is not well-posed: a very large number of images are compatible with the data. This is true in radio-interferometry in general, as it begins with a discrete sampling of the source, and fills in the gaps during image reconstruction. But the lack of well-posedness is exacerbated in cases where the data are sparse and noisy.

First, the sampling of the $(u,v)$ plane is limited to a discrete set of points. For $N$ telescopes one has up to $N(N-1)/2$ non-overlapping baselines, which gives $N(N-1)/2$ points in the $(u, v)$ plane. Interferometry on short wavelengths requires long baselines. In practice, this means that few components are available to form an ad hoc array and the collected fringe visibilities in short millimeter wavelengths are sparse. Obtaining a less sparse sampling is possible when Earth's rotation is taken into account; this ``sweeping'' effect can be clearly seen in the ellipsoidal patterns of, \textit{e.g.}, figure 2 of \textcite{2019EHT_M87_paper1}. From a technical standpoint, interpolation techniques (such as deconvolution used in the CLEAN-type algorithms), which fill the gaps by allowing ascription of non-zero values for the unmeasured components of visibility, can be used to partly remedy this first problem. However, philosophical concerns about underdetermination cannot be overcome by such ascriptions, since these remain \textit{unmeasured} components that could in principle take on multiple different values.

Second, visibility measurements are influenced by various types of noise, which further corrupt already sparsely sampled information. These include scattering due to propagation of radio waves in the solar wind and in the Earth's ionosphere; atmospheric noise (for instance phase fluctuations due to variable water vapor distribution in the troposphere); noise internal to components of an antenna (including, for small wavelengths employed in VLBI, both quantum and thermal noise); and, for sub-mm wavelengths, interstellar scattering \autocite{doeleman2008event}.

Producing an image from such data amounts to a judgment of which of these images is most likely---a judgment that depends on the background assumptions and priors that one builds into the imaging process. This is an important concern, because one might well worry that theoretical assumptions or practitioners' biases may be built into an interpretation of the collected data. The sparse and noisy data obtained with VLBI lead to general problems concerning underdetermination and theory-ladenness, presenting a challenge for assessing the results of VLBI observations.\footnote{Theory ladenness of, among other things, EHT parameter extraction is studied separately in [BLINDED FOR REVIEW].} The concern here is that the data may provide insufficient information about the source emission (due to sparseness and noisiness) to determine source features, and so astronomers will build assumptions into the imaging process that all but guarantee the image has certain expected features (e.g., ring-like, with a central brightness depression). If so, then the final image would reflect bias in the imaging process, rather than what the data are revealing about the source.

The challenge is especially pressing in the case of EHT for three key reasons. First, there are fewer sites in the array than for other VLBI measurements, increasing the sparseness of the data \autocite[9]{2019EHT_M87_paper4}. Second, the experiment is (in practice) unique. There are no previous 1.3mm observations of the EHT sources and no comparable black hole images on the event-horizon scale.\footnote{Strictly speaking, the event horizon is expected to be located within 2/3 of the observed bright asymmetric ring interpreted as the black hole shadow. See also footnote 1 of \textcite{bronzwaer2021nature} for a discussion of black hole shadows and some related terms such as photon ring and central brightness depression.} There are also no operating alternative arrays capable of replicating the EHT's observations. Third, insofar as the EHT results are used to probe a previously unexplored physical regime, it is particularly important that theoretical assumptions are not baked into the imaging procedures, since doing so could risk a vicious circularity. An additional factor is that VLBI takes place within the epistemic context of astrophysics, which some see as methodologically impoverished relative to more traditional experimental sciences (we briefly discuss this below in section \ref{astrophysics}).

Due to underdetermination of the image features, and sparseness and noisiness of the EHT data, additional assumptions have to be brought in to the imaging process. What needs to be shown is that the features of the final image do not depend on the specific assumptions built in, but rather seem to be favoured by the data themselves independently from any particular set of imaging assumptions.

So, given these challenges, why should one trust the result of the EHT observations?

Our answer is that confidence in the result should be understood as being based on the robustness of key features of the image.\footnote{Of course, there are many other factors influencing the assessment of the final result, most importantly: confidence that the operational equipment functions correctly, that the data have been correctly calibrated, and that the noise has been properly subtracted during data reduction. We will not be discussing this in detail here; partly due to space constraints of this paper, and partly because these issues are not specific to large experiments in our sense.} Although an infinite number of images are compatible with the EHT data, certain features appear to be stable across a range of approaches to producing images from this data. This is compatible with how the EHT collaboration itself justifies the final image:  
\begin{quote}
A number of elements reinforce the robustness of our image and the conclusion that it is consistent with the shadow of a black hole as predicted by GR. [...] [O]ur analysis has used multiple independent calibration and imaging techniques, as well as four independent data sets taken on four different days in two separate frequency bands. \autocite[section 8]{2019EHT_M87_paper1}\end{quote}
Two questions arise. First: what argumentative strategy is being used? In particular, does it indeed take the shape of a robustness argument? (In section \ref{EHT_2019} we reply affirmatively to that.) Second: is there an analysis of robustness available which accounts for the correctness of said argumentative strategy? (Again, the answer of section \ref{conclusions} is affirmative.)
In order to answer these questions, we now turn to examining accounts of robustness that have emerged within the philosophy of science literature. 

\section{The idea of robustness}\label{robustness}
Philosophers of science have devoted much attention to the idea of \emph{robustness}, where robust results are, at first pass, convergent results derived from a variety of derivations, tests, or lines of evidence.\footnote{There is a long and rich tradition of philosophical interest in ``robustness'' (including related ideas such as ``consilience'', ``stability'', and ``reproducibility''), which we cannot discuss in detail here. \textcite{Wimsatt1981} provides a detailed philosophical discussion of ``robustness analysis'', drawing on the earlier work of \textcite{Campbell1958}, \textcite{Campbell-Fiske1959}, \textcite{Levins1966}. Wimsatt also points to \citeauthor{Peirce1868} [1868] (1936) and to Whewell's ``consilience of inductions'' (via \textcite{Laudan1971}) as even earlier philosophical precursors. \textcite{Woodward2006} provides another influential analysis of robustness. For related discussions of the stability of phenomena across variations in experimental interventions see \textit{e.g.}, \textcite{Hacking1983,Franklin-Howson1984,Galison1987}, and for recent discussions in the context of the LHC, see \textcite{karaca2020two} and \textcite{bogerobustnes}.} The basic idea is that the convergence of results based on varied evidence provides extra reason to trust the shared conclusion. Such a conclusion is said to be ``robust'' compared to a conclusion based on a single line of evidence.

Under the right conditions (to be discussed below), convergent results improve the security of an inference because having independent lines of evidence pointing to the same conclusion bolsters the inference against ``error scenarios''---ways that faulty assumptions could be leading the inference awry.
A classic and often-cited example is the agreement across a variety of measurements of Avogadro's number, which \textcite{Perrin1913} used to argue for the reality of atoms \autocite{Salmon1984,Cartwright1991,Mayo1996,schupbach2018robustness,Dethier2020}. A similarly prominent, though more controversial, example is the use of agreement across climate models to bolster confidence in the results obtained from these models (see, for example, \textcite{Lloyd2010,Parker2011,Winsberg2018,Dethier2020}).\footnote{However, see \textcite{Dethier2022Unity} for discussion of why these paradigmatic examples may be misleading: roughly, pointing to an exemplary case of robustness analysis in an experimental context (Perrin) and a particularly challenging case for robustness analysis in a modeling context (climate models) biases us against acknowledging the potential confirmatory power of robustness analyses in the latter context.}

\subsection{Robustness as variation in auxiliary assumptions}\label{auxiliary_conditions}

Establishing the robustness of a result will generally involve a careful examination of the extent to which the evidence is genuinely varied. If an ``evidence claim'' (of the form ``$E$ is evidence for $C$'') depends on a set of auxiliary assumptions ($a_{1}, \ldots, a_{n}$), then the fallibility of the auxiliary assumptions can lead to the failure of the evidence claim. 
For the convergence of evidence itself to be evidence that the conclusion is true, it is necessary to eliminate factors that could lead to convergence on a false conclusion. For example, if $E_{1}$ and $E_{2}$ only count as evidence for $C$ in virtue of some shared auxiliary assumption, then it is possible that the failure of this assumption is responsible for the convergence. If two tests of a hypothesis rely on similar sets of assumptions (perhaps because two experiments are conducted under similar conditions) then the evidence provided will not be very varied and agreement across these tests provides only minimal increase in confidence in $C$.

\textcite{Staley2004robustness} argues that there are several ways that robustness can make an evidence claim more secure. He distinguishes between the \emph{strength} and \emph{security} of an evidence claim. The \emph{strength} of an evidence claim concerns the degree to which the data supports a given hypothesis. Any such claim will appeal to a number of auxiliary assumptions---assumptions that are needed to support an evidence claim in the context of a particular experiment. The \emph{security} of an evidence claim then concerns ``the degree to which the claim that some result is evidence for a hypothesis is itself susceptible to defeat from the failure of an auxiliary assumption'' \autocite[470]{Staley2004robustness}. In other words, security concerns the likelihood that a given evidence claim fails due to the failure of an assumption used in making that claim. Staley then argues that robustness improves the security of an evidence claim.

A similar analysis of robustness is given by \textcite{Dethier2020}, who defines robustness as agreement across varied sources of evidence, where ``variation in evidence'' is understood as variation in ``auxiliary conditions''---the `physical conditions and theoretical facts that must obtain' in order for a test to provide evidence for a hypothesis \autocite[74-5]{Dethier2020}.\footnote{Since the focus on this paper is on inferences made during the imaging process, we prefer to use the term auxiliary assumptions; but we take those assumptions to express what Dethier calls auxiliary conditions.} Note that Dethier explicitly defends a unified analysis of robustness across appropriately varied models and experiments---a position he calls `unity' \autocite{Dethier2022Unity}.

Convergence only constitutes robustness insofar as the evidence is varied. That is to say, the robustness of the results depends on the degree to which they are stable across a range of auxiliary conditions. However, there are also cases where variation does little to address the overall fallibility of the auxiliary assumptions. \textcite[section 5]{Staley2004robustness} discusses two such cases: \textit{spurious convergence} and \textit{failures of independence}.

Spurious convergence occurs when one test is likely to produce a particular result 'regardless of whether or not the assumption in question is true' \autocite[9]{Staley2004robustness}. In other words, one test fails to provide reliable evidence for the phenomenon.
For example, if one of the EHT's imaging pipelines (discussed in detail in section \ref{EHT_2019} below) is likely to produce an image like the 2019 image \textit{even for very different data} then agreement between this and another imaging pipeline does not add to the robustness of the result. In such cases, convergence between two tests is meaningless and does not increase the security of the overall evidence claim. 

Failures of independence occur where tests are less independent than previously thought, in that overlapping assumptions turn out to be responsible for the convergence. (An instructive recent example of this is described in \textcite{Gueguen2020}, which we discuss in more detail in the final section \ref{conclusions}.)

In addition to robustness, or \emph{convergent validation}, Staley argues for the importance of a further requirement, namely \emph{discriminant validation}: the demand that different sources of evidence do not yield convergent results when the phenomenon being measured is absent. For the EHT, one way to spell out this requirement would be as follows: the methods adopted by different imaging pipelines ought to produce distinct results if applied to data sets from different sources. Convergence should only occur where the convergence can be explained by similarities in the data set and the source systems that produced it. 

For the purposes of this paper, another class of cases discussed by \textcite{Staley2020} is of particular interest: situations in which the models used to make inferences (within the context of a particular experiment) are varied in order to account for systematic uncertainty. This can reflect the experimenter's ignorance about which model best represents the actual situation. Staley argues that the estimation of systematic bias can be understood as a kind of robustness analysis. In particular, altering the model used in a measurement (understood as a model-based inference) leads to an ensemble of ``measurements'' using different modeling assumptions. Based on this ensemble, experimenters employ what Staley calls a weakening strategy: they can report a ``weakened'' conclusion, secured by consideration of the experimenter's imperfect knowledge (\textit{i.e.}, by the robustness analysis).\footnote{In a similar vein, \textcite{ritsonstaley2021uncertainty} (following \textcite{beauchemin2017autopsy}) discuss weakening of conclusions (understood as addition of larger error bars) as a strategy of mitigating reliance on theoretical assumptions, and thus as a way of remedying circularity.}

In section \ref{EHT_2019}, we argue that the use of multiple imaging pipelines by the EHT (especially the way that they are used together to produce the final image) is an instance of a similar weakening strategy. However (and in contrast to some claims, such as \textcite{weinstein2020coincidence}), the imaging algorithms being used are not \textit{models} that represent the target system. 
At the same time, variation in these algorithms is intuitively not variation in the \textit{experiment}, at least not in the way the experiments were varied in (for example) the different detections of Brownian motion or the measurements of Avogadro's number (see \textcite{schupbach2018robustness} for discussion of these and other examples). For a variation in algorithms is not a variation in initial conditions, population, or observational setup.
Nevertheless, we argue that the arguments put forward in the EHT papers should be understood as robustness analyses, though these analyses occupy a kind of middle ground between variation in experimental conditions and variation in models. 

\subsection{The epistemic value of varied models}\label{varied_models}

Most philosophers endorse the epistemic value of robustness with respect to variation in experimental conditions.\footnote{A notable exception is \textcite{Hudson2014}.} There is less consensus when it comes to the value of robustness across variation in models. For example, \textcite{Cartwright1991}, \textcite{Orzack-Sober1993} and \textcite{Woodward2006} argue against the evidential value of robustness across varied models, while  \textcite{Weisberg2006,schupbach2018robustness,Dethier2020,Staley2004robustness} argue in its favor.
 
\textcite{Cartwright1991} claims that variation in models does not confer the same epistemic benefits as variation in experimental methods or instruments. In the latter case, Cartwright thinks that the variation allows for an ``argument from coincidence''; while any individual experiment might be based on faulty assumptions or equipment, 'it would really be a coincidence if each of the separate methods independently produced the same wrong result' \autocite[150]{Cartwright1991}. However, Cartwright doesn't think that this reasoning works for varied models because these:
\begin{quote}
        do not constitute independent instruments doing different things, but rather different ways of doing the same thing: instead of being unrelated, they are often \textit{alternatives} to one another, sometimes even contradictory [\ldots] [W]e look at the phenomenon with at the very most one instrument which could be operating properly. \autocite[153-4]{Cartwright1991}
    \end{quote}
The idea here is that \textit{at most} one model is correct or ``operating properly'' while any alternatives are faulty. Thus varied models cannot ground arguments from coincidence like varied experimental methods can. \textcite{Orzack-Sober1993} also take issue with robustness analyses of this kind for essentially the same reason. They argue that the joint prediction of a particular result by a set of models can only be relied on if we know that one of those models is true \autocite[538-9]{Orzack-Sober1993}.

The core assumption here seems to be that false models cannot make reliable predictions. Since robustness analyses involve finding truth at 'the intersection of independent lies' \autocite[423]{Levins1966}, this leads to rejection of the supposed epistemic value of robustness across variation in models. 

There are some good reasons to reject these assumptions about the evidential value of false models. Recent literature concerning scientific modeling emphasizes that models are tools, made and used with a particular epistemic or practical purpose and judged by their adequacy-for-purpose rather than their truth or representational accuracy (see \textit{e.g.}, \textcite{Morgan-Morrison1999,Giere2010,Parker2010,Parker2020,Currie2017}). \textcite[31]{Bokulich-Parker2021} call this the ``pragmatic turn'' in the philosophical treatment of scientific modeling.

\textcite{Dethier2020} draws on work in this ``pragmatic turn'' to argue that arguments such as Cartwright's are based on a mistaken emphasis on the role of models in representing a particular target system, rather than on their contextual role within a broader empirical investigation. Dethier argues that models are tools that are either adequate or inadequate to use in a particular context. This recovers the analogy with variation in experimental instruments, and leads to a unified view of robustness across experiments and models. 
\textcite{Weisberg2006} has also specifically responded to \textcite{Orzack-Sober1993} by pointing to the low-level empirical confirmation that supports the use of a particular modeling framework in the first place. His response is thus similarly based on an account of robustness analysis that places it within the context of a broader empirical investigation. 

\textcite{weinstein2020coincidence} has recently applied a version of Cartwright's argument in the context of the EHT, arguing that the agreement across images fails to provide the basis for an argument for the reliability of these images. Drawing on our own analysis of the arguments provided by the EHT Collaboration in section \ref{EHT_2019}, we respond to this version of the argument in section \ref{conclusions}. 

\subsection{Robustness in astrophysics}\label{astrophysics}

Establishing robustness may not be necessary where confidence in all of the auxiliary assumptions is high. Thus \textcite[150-1]{Cartwright1991} acknowledges that \textit{reproducibility} is not necessary for (ordinary, lab-based) experiments where we already have high confidence in our instruments. When the entire experiment plays out in a lab, it is often possible to independently check the validity of auxiliary assumptions through, for example, extensive calibration procedures. However, robustness analyses are of particular significance in contexts where checks on auxiliary assumptions---specifically those that are plausible sources of error---are difficult or impossible to perform. 

Due to the lack of controlled experiments, model-dependence of observations, and long characteristic timescales of target systems evolution, astrophysics is sometimes thought to be methodologically impoverished compared to other sciences.\footnote{Philosophers such as \textcite{Anderl2016} and \textcite{Jacquart2020} have discussed the distinctive epistemic challenges (and methods) of astrophysics. \textcite{Anderl2016} and \textcite[ch.4]{Elder2020} also draw connections between astrophysics and historical sciences, which (like astrophysics) are sometimes thought to be methodologically impoverished compared to experimental sciences. See, \textit{e.g.}, \textcite{Cleland2002,Currie2018} for discussion of the methodology of historical sciences.} Astrophysical systems also usually have at least some extreme properties that can not be replicated on Earth. These features lead to some special challenges for validating auxiliary assumptions about astrophysical target systems. Without controlled experiments, astrophysicists must reason backwards from causally downstream observations of distant target systems. Any such inference depends on a range of auxiliary assumptions about the source, the transmission process, and the detector \autocite{Shapere1982}. Many of these assumptions about the target cannot be independently checked, nor can the relevant conditions be replicated in terrestrial labs.\footnote{As \textcite[654]{Anderl2016} puts it: `[a]strophysics deals with phenomena and processes that are found occurring in significantly more extreme conditions than anything that can be artificially generated in a terrestrial laboratory. The range of temperatures, pressures, spatial scales, and time scales pertinent to astrophysical phenomena are larger than anything that is commonly accessible to humans by direct experience. Also, the often dominant influence of gravity sets astrophysical processes apart from terrestrial laboratory settings.' In more concrete terms, in the ongoing observational run the Large Hadron Collider reaches collision energy of 13.6 TeV. But, in a particularly striking example, a particular subtype of AGNs, PeVatron blazars, accelerate particles to PeV energies (with 13.6 TeV equal to 0.0136 PeV). In terms of mass, M87* is approximately $6 \times 10^{9}$ solar masses; just in terms of the spatial area it takes on the sky, the whole Solar System up to Pluto's orbit would easily fit inside its diameter, with some room to spare.} It is thus epistemically prudent to weaken the dependence of the results on particular auxiliary assumptions through robustness analyses. 

In a unique experiment such as the EHT, robustness might initially seem out of reach. The EHT array is the only one capable of collecting the relevant data (\textit{i.e.}, data pertaining to the near horizon structure of a supermassive black hole), because no alternative array reaches the required resolution.\footnote{Naturally, the array could be extended, for example with an orbiting component. Such extensions are likely to include some form of the EHT array as a proper subset, and since their construction and use will be built on the EHT array success, understanding sources of robustness of the EHT results is of importance for this future work. Note, though, that the EHT extended with an orbiting component will be less troubled by issues arising from the sparse sampling, but more impacted by thermal noise for space-ground baselines; see figure 4 of \textcite{palumbo2019metrics}.} 
Moreover, in 2017 the primary EHT targets included just two systems: M87* and SgrA*.\footnote{There are also the secondary targets, some of which do not involve observations of black hole shadows; for instance the jet structure of the 3C 279 blazar has been investigated on the basis of 2017 data \autocite{2020EHT3C279}. See also footnote \ref{fn:calibration_sources}, above.} Varying the experimental setup or the population to get varied evidence is simply not a viable option (the latter is not available, whereas the first would lead to loss in resolution). The data gathered by the EHT are essentially a unique line of evidence about the horizon-scale structure of a supermassive black hole. 
Thus robustness cannot be established on the grounds of variation in the physical processes involved in data \emph{collection}. 
Such experiments face a clear threat of model-dependence (or theory-ladenness) and underdetermination, since the usual methods for varying evidence about the target are blocked.
In such situations, robustness becomes important during the data \emph{analysis}.
As we will discuss in section \ref{EHT_2019}, the overall approach to data analysis leading to the 2019 EHT image reflects this need to ensure that results are robust despite the lack of independent lines of evidence about the target system. 

The example of the infamous BICEP2 retraction provides a useful illustration of the importance of establishing the robustness of results in astrophysics.\footnote{\textcite{Keating2018LtNP} provides a detailed account of this saga. See also \textcite{BICEP2014} for scientific details.}
The BICEP experiment initially claimed to have detected gravitational waves, through the detection of a polarization pattern in the cosmic microwave background (CMB). This pattern was attributed to B-modes of primordial gravitational waves, and consequently, was interpreted as evidence for inflation. B-modes observed by BICEP2 can be generated in many ways, primordial gravitational waves merely one among many, so effects of (a) lensing B-modes of distant galaxies and (b) dust B-modes had to be subtracted from the overall signal. BICEP2 modeling of (b) turned out to be inadequate, in that (b) could account for the entire signal that had been attributed as the gravitational wave contribution. So the BICEP2 claim was not sufficiently robust across variation in modeling of dust, and consequently not secure against a failure of an auxiliary assumption: the particular model of dust-induced polarization. Fallibility of this auxiliary assumption undermined the final result.
But there is a second lesson to be drawn. In the BICEP2 case, Planck data from the same year showed that the model used by BICEP2 was wrong. In the absence of such data it may have been possible for the flawed claim to persist unchecked for a long period of time. This illustrates the importance of demonstrating the robustness of a result across variation in auxiliary assumptions in case one's model is wrong (as BICEP2's turned out to be). This is particularly essential in cases where there is no alternative data collection method at hand (or no data able to confirm auxiliary assumptions).\footnote{Attempts at detecting gravitational waves using Weber's bar experiments are another episode which can be interpreted as lack of sufficient demonstration of the robustness, this time in the experimental setup. In that case, inability to reproduce the result led to the dismissal of Weber's discovery claim. A history of those attempts can be found in \textcite{collins2010gravity}.}

\section{The 2019 Event Horizon Telescope image}\label{EHT_2019}

As we have pointed out in part \ref{interferometry_VLBI}, some form of robustness plays an explicit role in the justification provided by the EHT Collaboration for taking their final image to be a faithful representation of M87*. Throughout the various processes involved in the analysis of the 2017 observational data, the convergence of results across variation in auxiliary assumptions is used to validate those results. This includes variation in the moment of data collection, calibration methods, imaging algorithms, and imaging teams. In this section, we will discuss the robustness of the 2019 images across such variation, with a particular focus on the imaging algorithms and teams; in section \ref{conclusions} we will then relate this argumentative strategy to some recent accounts of robustness. Due to space constraints, in this paper we do not discuss robustness and the roles various types of models (GRMHD simulations) play in the source modeling and extraction of physical parameters such as the mass of the central object (particularly in \textcite{2019EHT_M87_paper5,2019EHT_M87_paper6}).

Before going into details, we note that there is also a ``historical'' aspect to the robustness of the EHT image of M87*, having to do with the understanding of the reliable functioning of the array itself. Recall that the 2017 observations (on which the 2019 results are based) were made with an array consisting of eight elements, seven of which were used for the M87* target. However, pre-EHT arrays consisted of only three stations in 2009, with successively larger arrays observing the primary EHT targets over time; these arrays were too sparse to resolve M87*.\footnote{Recall, from section \ref{interferometry_VLBI}, that more stations give more non-overlapping baselines, which means better resolving capability, because the source will be sampled at a larger number of points. See Figure 2 and Table 1 in \textcite{wielgus2020monitoring} for a summary of the size of pre-EHT arrays between 2009 and 2017 and comparison with the 2017 EHT array. Figure 1 of \textcite{ngeht-key-science-goals} shows the density of the planned ngEHT array, while Figure 3 of \textcite{palumbo2019metrics} shows the gains in baseline coverage that might be obtained with the use of an orbiting component.} The consistency (and progressive increase in resolution) of the results as the array size increased over time provided confidence in the reliability of the 2017 observations at the level of the data collection.\footnote{\textcite{kennefick2019no} has a related discussion of the case of the tests of general relativity by the 1919 eclipse expedition. Observations subsequent to the 1919 expedition failed to improve upon the precision of the 1919 results, which cast doubt on their reliability.} \footnote{The process of establishing trust in the array itself might be also accounted for in terms of Guralp's notion of endorsement. Paraphrasing Guralp, a measurement scheme is said to be endorsed with respect to some target phenomenon when that target is considered to be epistemically accessible within the current account of said measurement scheme; see \textcite{guralp2020evidence} for an illustration of endorsement in the context of supernovae cosmology as used in the measurements of the accelerating expansion of the universe. Guralp's analysis of endorsement and robustness is embedded within an iterative model of knowledge production due to \textcite{chang2004inventing}. As such it is useful for understanding the claim that an EHT-type array is a reliable scientific instrument, but its' usefulness for understanding claims of robustness of a particular result based on a particular dataset, such as the M87* image, is limited.} 

We will now provide an overview of the two main approaches to image reconstruction (together with a description of the parameters EHT used), and then discuss main stages of the EHT's imaging process.

\subsection{Two approaches to imaging}\label{clean_rml}

Astronomers (including \textcite{2019EHT_M87_paper4}) often distinguish between inverse modeling and forward modeling. Inverse modeling begins with an inverse Fourier transform of the data (in VLBI: sampled visibilities), and proceeds by correcting for artifacts of the instrument. In contrast, forward modeling begins with an image, and uses its Fourier transform to evaluate whether it is consistent with the data. CLEAN is an instance of inverse modeling, and two algorithms developed for the EHT, eht-imaging and SMILI, are instances of forward modeling (methods based on similar principles have been used elsewhere in radio astronomy).

\subsubsection{The standard approach: CLEAN}\label{CLEAN}
Broadly speaking, CLEAN is the standard approach to interpretation of data in radio astronomy (for a detailed discussion of the CLEAN method, see \textcite[ch.11]{Thompsonetal2017}). CLEAN is based on the assumption that the image consists of a number of point sources. It is an iterative procedure in which (speaking loosely) the brightest regions are subtracted from the initial ``dirty'' image and added as delta functions to the ``clean'' image. This is repeated until all points with intensity above a certain brightness threshold have been removed, so that any remaining brightness in the image comes from residuals below it. Next, the delta functions that comprise the clean image are convolved with a clean beam function (which is typically a Gaussian distribution centered around some function of the initially subtracted bright region), and residuals are added. Groups of delta functions may represent extended structure, so CLEAN interpolates between these points.

A general problem with CLEAN algorithms is the possibility of emergence of spurious structures located at intervals equal to the spacing between subtracted regions. CLEAN is also non-linear, because the deconvolution procedure is non-linear. Sufficient conditions under which its successive applications converge are known; this in particular requires that CLEAN is applied to only a limited area (called a ``window'') of the original image.

There are many variants of CLEAN (including Clark's algorithm, the Cotton-Schwab algorithm, multiscale CLEAN, and others); the EHT used DIFMAP, a scripted version of CLEAN \autocite{2019EHT_M87_paper4}. CLEAN was performed manually during the first stage of the EHT imaging process, but a scripted version was used in the second stage. The user input has five parameters: total compact flux density (in other words, how bright is the central region which is being imaged); a condition determining when the iterative procedure stops (two such conditions were used: when total flux has been reached, and when the image became too noisy); the diameter of the region under consideration (the window); down-weighting of some data (baselines using ALMA have higher signal to noise ratio, so without down-weighting these baselines would dominate the image); and a parameter further weighting some data by the visibility errors.

\subsubsection{Regularized Maximum Likelihood methods}\label{RML}

RML methods search for an image which minimizes a specified objective function. Schematically, the objective function takes the form: (sum over weighted data terms) minus (sum over weighted regularizers). These weights are often called hyperparameters, and are used to balance between fit with the data and other preferred features of an image (for example, its overall continuity). The objective function is often interpreted as a log-likelihood of a posterior probability, \textit{i.e.,} the goodness of fit of a particular image (or, in general, of a model) given some new information.

The user input for each RML method consists of the specification of an objective function, that is, a decision about what regularizers to use, what
values to assign to parameters for the regularizer terms, and what weights to use. The EHT used a total of six different regularizers: both eht-imaging and SMILI include total compact flux density, total variation TV, and total squared variation TSV; eht-imaging further included MEM and $l_{1}$, and SMILI included weighted $l_{1}$. Total compact flux density requires that reconstructed images have energy flux density near a specified value in Jansky; a value of 0.66 is considered the most plausible on physical grounds, but the range of 0.4 to 0.8 has been explored. Other regularization terms encode broad assumptions about which features of an image are favored. MEM is a relative entropy measure which prefers similarity to a prior image, which was assumed to be a circular Gaussian image with variable Full Width Half Maximum (FWHM; this is a common measure of quality of an image, describing how much a point of light is smeared out). $l_{1}$ and weighted $l_{1}$ favor sparse images, TV favors piecewise smooth images with regions separated by sharp edges, whereas TSV favors overall smooth images. See appendix A of \textcite{2019EHT_M87_paper4} for the detailed description of the regularizers, and table 3 on p. 17 for the values used in the final imaging (these values are determined by the parameter surveys performed in the second stage described below). In cases of shared regularizers, the numerical values of final (fiducial) parameters differ to some extent between eht-imaging and SMILI.

We note that the above description strongly suggests that the two RML pipelines do not encode the same set of user assumptions, but rather can be thought of as independent implementations of the same broad approach to imaging. Furthermore, these methods are largely independent of the CLEAN method discussed in section \ref{CLEAN}; the only assumption that all three imaging algorithms share is the physical assumption about the total compact flux of the region (limited to total of 100 $\mu$as, and measured at 230 GHz). This value is estimated using both the EHT data and with other VLBI data sets from around the same time observing at longer wavelengths; appendix B of \textcite{2019EHT_M87_paper4} describes this in detail.

\subsection{The EHT imaging process}

For our purposes, the upshot of the discussion of imaging algorithms is that although some user input is necessary for any of the imaging methods used by the EHT, the implementation and interpretation of said input differs significantly between these methods. As discussed earlier, some additional assumptions are needed in order to be able to solve the imaging inverse problem, which otherwise is not well-posed. But during the later stages of the analysis it will turn out that the features of the final image do not depend on any particular set of those assumptions. This will also be of relevance in section \ref{conclusions}, where we will discuss concerns that shared numerical artifacts might be responsible for convergence of results.

The production of images representing M87* took place in two main stages. In the first stage, four imaging teams worked independently on a subset of data, without any communication between them, each to produce an initial image. In the second stage, the range of parameters for which these algorithms produced convergent results was explored. The final imaging was conducted using three different algorithms and parameters determined in the second stage, and included an averaging process.

\subsubsection{First stage: blind imaging}\label{stage_I}
The first imaging stage started with the creation of four separate imaging teams, the members of which were forbidden from communicating with members of other teams.\footnote{One might also consider further blinding, for example by feeding one of the teams entirely synthetic data. However, trained astronomers could easily see some general features of the data (like the ring) already from calibrated visibility amplitudes (figure 2 in \textcite{2019EHT_M87_paper1}), so this might be seen as an unnecessary use of scientists' time.} Although there were no restrictions on what data-processing and imaging procedures each team used, they happened to divide up evenly between two main methods: CLEAN and RML (these methods are discussed above). In particular, teams 1 and 2 primarily used RML and teams 3 and 4 primarily used CLEAN. The initial imaging results were based on an early-release engineering data set from observations on April 11, 2017.\footnote{This data set was not fully calibrated: it had both \textit{a priori} and network calibration, but lacked calibrated relative polarization gains (for discussion of the details of calibration, see \textcite{2019EHT_M87_paper3}). The data collected at two frequency bands were correlated separately: "the Haystack correlator handled the low-frequency band (centered at 227.1 GHz), with MPIfR correlating the high band (centered at 229.1 GHz)", \textcite[4]{2019EHT_M87_paper3}. This provides an additional safeguard against the worry that there is either some systematic bias or a dominant noise contribution in any given frequency band, or, of course, a calibration error.}

Both of the imaging methods---CLEAN and RML--- are sensitive to various choices made in the process and thus potentially susceptible to being influenced by biases of the user, making it ``difficult to assess what image properties are reliable from a given imaging method'' \autocite[9]{2019EHT_M87_paper4}. The level of user input required to apply these methods led to particular concerns about false confidence and collective confirmation bias. While these are general concerns for VLBI imaging, they are particularly acute for the EHT due to both the sparseness of the sampling and the uniqueness of the measurement (as discussed at the end of section \ref{epistemic_situation_large_experiments}). The risk of collective confirmation bias is explicitly cited by the EHT as the key reason for dividing up into independent imaging teams for the first stage \autocite[9]{2019EHT_M87_paper4}.\footnote{Another example of this strategy is found in the LHC. Here there are two independent lines of evidence in virtue of being split into two independent groups, ATLAS and CMS, doing both the data collection and interpretation separately. However, there is also a dis-analogy with the EHT case since the two groups use two differently designed detectors. Arguably this provides a stronger case for robustness of the results, because the variation in the instruments provides an independent stream of evidence about the phenomenon being probed, rather than merely an independent analysis of a single data set. The two independent (modeled) search pipelines (GstLAL and PyCBC) in the LIGO-Virgo experiments play a similar role \autocite{LVC2016search}.}\footnote{The number of independent lines of investigation, the extent to which these are blinded, and whether blinding will be at all present in the future is highly contingent. Once sufficient trust in the data analysis pipelines is established, it might be expected that this feature of the methodology may be limited or even eliminated in future observations, particularly with the same resolution.}
Features of the image that are robust across pipelines are less likely to be products of such confirmation bias, given the independence of the teams. 

Comparisons of the results of different pipelines revealed that some features of the images were consistent across the four pipelines, while others varied substantially. All four images were dominated by a ring with a diameter of 40$\mu$as, with brighter emission to the south. These features of the image are \emph{robust} across variation in the imaging teams and the methods they employed. 
However, further features of the image---including the ring's azimuthal profile, thickness, and brightness---were not consistent across the images. As \textcite[9]{2019EHT_M87_paper4} notes:
\begin{quote}
The initial blind imaging stage indicated that the image of M87 is dominated by a $\sim40$ $\mu$as ring. The ring persists across the imaging methods.
\end{quote}
Thus we see that the robust features of the images---but not others---are trusted as faithful representations of M87*.

If robustness concerns varying assumptions of an inference, what about these assumptions has been varied so far? To some extent, imaging methods. However, systematic variation across imaging algorithms was the main focus of the second stage of the imaging process. The main target of the first stage is largely social: the avoidance of errors due to the choices and expectations of the particular humans doing the imaging. The value of this  kind of robustness is limited if it turns out that there is some reason to think that the biases of the different groups are not appropriately independent. For example, if there is a significant overlap in the education and career paths across members of different imaging teams, then there may be some degree of pre-coordination of the choices they are likely to make using CLEAN and RML techniques. Nonetheless, the approach taken in stage one does at least some important work demonstrating how different choices made by epistemic peers can result in differences in certain features of the images. Other more robust features can be seen as reflecting a kind of consensus.

\subsubsection{Second stage: parameter surveys}\label{stage_II}
As just discussed, each of DIFMAP, eht-imaging, and SMILI requires selecting some parameter values in the implementation of the algorithm. The goal of the second stage was to 'explore the dependence of the reconstructed images on imaging assumptions' \textcite[9]{2019EHT_M87_paper4}: instead of simply basing this selection on expert choices, for each algorithm the EHT collaboration performed a scripted parameter survey to determine the appropriate inputs. 

This process started with the production of synthetic data. Those data were used to determine appropriate parameters through parameter surveys; in turn, these parameters were further used in the final imaging. (These steps correspond to sections 6.1, 6.3, and 7.1 respectively of \textcite{2019EHT_M87_paper4}).) Finally (as discussed in the next section \ref{imaging_process}), the outcomes of this imaging process for the three different pipelines were compared and combined in order to draw reliable conclusions about the features of M87*.

First, four synthetic data were generated by starting with model images (``ground truth images'') then synthesizing the data that the EHT would have recorded for such objects (using the eht-imaging software). Four simple geometric models with different morphologies were chosen for this task, each with visibility amplitudes similar to those observed in M87* \autocite[10]{2019EHT_M87_paper4}. These models included: a ring, a crescent, a disk, and a double Gaussian (\textit{i.e.}, two separated Gaussian components). It is worth noting that none of these model images replicates all of the features of the EHT data. As such, they are not good candidates for final images of M87*. But they produce similar-enough data to optimize the imaging algorithms without over-tuning to the expected morphologies of realistic images of M87*. In particular, the GRMHD simulations are not used at this stage. Another VLBI data simulator, MeqSilhouette+rPICARD, was used as a cross-check; this simulator includes corruption of the data (with ``measured weather parameters and antenna pointing-offsets'', thus providing a degree of safeguard against fine-tuning the data analysis to a particular form of noise present in the actual data.)

Second, each of the imaging algorithms was used to reconstruct images from these synthetic data sets for a range of combinations of parameter values (1008 for DIFMAP, 37500 for eht-imaging, and 10800 for SMILI). These reconstructed images were then compared to the original ground truth image used to produce that data set. Parameter combinations that resulted in reconstructed images deemed sufficiently similar to the ground truth images provide the range of parameters (the Top Sets of 30 for DIFMAP, 1572 for eht-imaging, and 529 for SMILI; for the ranges of parameter values and fractions these take in the Top Sets for each algorithm, see table 3 of \textcite{2019EHT_M87_paper4}) for which the algorithm performed well on EHT-like synthetic data.\footnote{Whether by doing this a sufficiently large parameter space has been explored should be answered on a case-by-case basis for each parameter. Assuming that the answer is positive, this contributes to the robustness of the final image. Conversely, if the range for some parameter has been underexplored, then the robustness of the final image might be limited. However, pre-2019 imaging challenges also included unusual sources, such as ``Frosty the Snowman'', which further increases the confidence that the imaging algorithms would succeed for highly unexpected sources.} Additionally, the unique set of fiducial parameters that jointly optimized the performance of the algorithm (\textit{i.e.}, those that produced the best reconstruction of the ground truth images) was selected; these values were then used in the final imaging. \textcite[10]{2019EHT_M87_paper4} stress:
\begin{quote}
Our surveys are coarse-grained and do not completely explore the choices in the imaging process. Nevertheless, they identify regions of imaging parameter space that consistently produce faithful image reconstructions on synthetic data, and they help us identify which features of our reconstructions are consistent and which features vary with specific parameter choices
\end{quote}
In other words, the parameter surveys explore the robustness of image features across variation in parameters (for each of the three algorithms). The selection of the fiducial parameters used for imaging M87* reflects the outcome of this robustness analysis.

Overall, the second stage of the imaging process demonstrates the robustness of at least some features of the images based on synthetic data across variation in both imaging methods (CLEAN and RML), the specific algorithms used to implement RML (eht-imaging and SMILI), as well as values for the parameters encoding remaining choices about the algorithm implementation. This stage builds confidence that the imaging algorithms are reliable tools for reconstructing the source emission based on sparse, noisy data and that the assumptions built into these algorithms are not biasing the results towards images with particular features (such as being ring-like, with a central brightness depression).

\subsubsection{The robustness of the final image}\label{imaging_process}

After determining the reliable parameter values (i.e. the fiducial parameters) in the scripted survey stage, all three algorithms were used to construct tentative ('fiducial') images of M87* from the EHT data. Four images (see Figure 11 of \textcite[18]{2019EHT_M87_paper4}) were produced from each pipeline---one for each day of observation (April 5, 6, 10, and 11). The images are all broadly consistent: they each feature an asymmetric ring with a diameter of approximately 40$\mu as$ and each of these rings is brighter in the south. Some other properties, such as the peak brightness temperature and the presence of a prominent depression in the ring's center, are also consistent across the images, but some differences remain. DIFMAP images have small central brightness depression and lower peak brightness temperature; RML images are more similar to each other than to DIFMAP, but SMILI ones are fainter inside and outside the ring than eht-imaging ones. This is because eht-imaging has stronger preference for sparsity due to $l_{1}$ regularizer.\footnote{It is a matter of debate whether or not the differences in the properties of recorded data across this period are attributable to changes in the source itself \autocite[21, Appendix E]{2019EHT_M87_paper4}.}

Further assessment of the compatibility of the images is achieved by blurring them each to obtain a 'common, conservative resolution' \textcite[20]{2019EHT_M87_paper4}. For each of the pipelines, the blurred images from the April 11 observations are shown in Figure 14 of \textcite[21]{2019EHT_M87_paper4} (compare with the pre-blurred images of Figure 11, and note the differences in both the resolution and brightness temperature). The CLEAN image (from DIFMAP) was already blurred prior to this step, through convolution with a 20$\mu as$ beam in the course of the image production. The two RML images (from eht-imaging and SMILI) were also blurred (with circular Gaussian convolution kernels), in order to match the lower resolution of the CLEAN image. As a result, the three blurred images look very similar and have consistent ring diameter and overall asymmetry.

Finally, for each day of observations, simple averages of the blurred images (shown in figure 15 of \textcite{2019EHT_M87_paper4} are taken as the 'conservative representation of [the EHT Collaboration's] final M87 imaging results' \autocite[21]{2019EHT_M87_paper4}. This averaging process has the effect of further emphasizing the features of the three images that are common to all, while de-emphasizing those that differ across pipelines. This way of reporting the EHT's result reflects the robustness analysis resulting from varying the imaging algorithms, and, as with the previous step, this averaging can be seen as a weakening strategy, sacrificing precision in favor of security.\footnote{\textcite[section 4.3]{bogerobustnes} discusses similar issues related to trade-offs between security and informativeness in the context of high energy physics.} 

Recall (from the section \ref{auxiliary_conditions}) that the general idea of the weakening strategy described by \textcite{Staley2020} is that varying the models used during data analysis results in an ensemble of measurements, analogous to varying the experimental apparatus. This can be used to account for systematic uncertainty resulting from the experimenter's ignorance about the optimal model for the task at hand. In the case of the EHT, this plays out in the way that the images from the different pipelines are compared and ultimately combined to produce a final ``weakened'' result. Some of the precision is sacrificed in order to bolster the security of the overall evidence claim: the blurring of the higher resolution images from the RML pipelines removes some of the more fine-grained structure present in these images. These cannot be corroborated by lower-resolution CLEAN images and are thus considered to be less trustworthy than features that are robust across all three. As with Staley's examples, this strategy reflects the experimenters' acknowledgment that none of the three pipelines are known to be the optimal one, in the sense that the results of one ought to be trusted above the results of the others. Blurring all of the images weakens the final conclusion, in the sense that the images lose resolution and (potentially informative) structure. However, this also increases the security of the results, resulting in images whose features can be considered more trustworthy than their unblurred predecessors. We can think of the blurring as analogous to adding larger error bars to a weakened conclusion---sacrificing the precision of the result to improve confidence in the fidelity of the final images. 

In addition to the convergent validation, the final imaging process helps provide discriminant validation. Recall that this means that the algorithms must \textit{not} yield convergent results when applied to different data sets. This condition is met by the use of the same algorithms to image a range of geometric models (disc, crescent, etc.) used in the parameter surveys during the second stage. When the algorithms are applied to the synthetic data sets from these models, they yield convergent results when applied to the same data, and divergent results otherwise (this can be seen in figure 10 of \textcite{2019EHT_M87_paper4}).\footnote{Note that this process involves what the EHT Collaboration call ``cross validation'', with the parameters used for imaging a particular data set being trained on a selection of data sets from other geometric models rather than the entire training set. This further helps them to 'verify that the training sets do not overly constrain the outcomes' \autocite[16]{2019EHT_M87_paper4}.}

Other sources of convergent validation include the use of these methods to produce images of more familiar astrophysical sources (\textit{e.g.}, 3C 279) and tests whether the image persists under site removal (in other words, providing some control over the problem of sparsity of the data) and calibration errors. Discriminant validation is also provided by earlier (pre-2019) imaging challenges used to vet algorithms on a wide range of synthetic data. One imaging challenge notably included reconstruction of images of Frosty the Snowman. This supports the counterfactual claim that \textit{if} the core of M87 looked rather different from expectations \textit{then} the EHT procedure would have produced a different image.\footnote{The following example deserves a separate conceptual study, but we note in passing that a contested detection claim of the photon ring in M87* could in part be seen as a failure of robustness. A particular general relativistic prediction concerns the inner structure of the bright ring seen on the M87* image: namely, that the bright ring contains a nested series of strongly lensed and successively sharper and thinner subrings, which are generated by photons orbiting the black hole more than once. Successive numbers n denote the number of half-orbits completed by photons around the black hole. Such rings provide a probe of the source's geometry, and up to n=3 might be detectable using interferometric methods (assuming that one station is located at the Lagrange point $L_{2}$); see \textcite{johnson2020universal} for a detailed discussion of these features. The recent claim of the detection of the n=1 due to \textcite{broderick2022photon} is based on different implementations of a particular imaging strategy called hybrid imaging. This method, however, is arguably \autocite{tiede2022measuring} prone to false positives, and an alternative method, geometric self-fits, cannot distinguish the n=0 and n=1 ring at the EHT baselines.}

Overall, the EHT imaging procedures establish both that their imaging methods produce convergent results when applied to the same data, and that these imaging methods produce divergent results when applied to different data sets. This provides the basis for an argument from the robustness of the images across imaging pipelines to the reliability of the EHT images of M87*.

\section{Discussion: robustness for the EHT}\label{conclusions}
The case for confidence in the EHT images of M87 rests on a series of robustness arguments. In the previous section, we focused on how the different imaging pipelines vary the groups of people doing the imaging, the algorithms used to do so, and the values of parameters chosen in the implementation of these algorithms. 

In doing so, we have provided an analysis of the robustness argument used by the EHT Collaboration. Our analysis underscores the importance of robustness arguments in establishing the reliability of results in astrophysics, as well as in other large experiments (as described in section \ref{epistemic_situation_large_experiments}). But from some points of view our conclusion may seem paradoxical: in cases where experiments themselves are not being varied, the nature and value of robustness is controversial. We will now address such objections, and discuss some of the ways in which robustness arguments used in the EHT differ from other forms of robustness considered in the recent philosophical literature.

The benefits of our analysis become apparent in the context of a recent discussion of the EHT image, where \textcite{weinstein2020coincidence} invokes (and seems to endorse) \textcite{Cartwright1991}'s arguments against the value of robustness across variation in models. 
Weinstein's analysis questions the reproducibility of the image. She says: 
\begin{quote}
    Since the results are very robust, says Cartwright, we think that there must be some truth in them. She disagrees with this claim. That is because ``at the very best one and only one of these assumptions can be right''. We may look at the images produced by the three imaging pipelines but pipeline number four may generate an image with no shadow at all. If ``God's function'' is this fourth pipeline, then the three other pipelines teach us nothing. \autocite[73]{weinstein2020coincidence}
\end{quote}
The argument here seems to be that the imaging algorithms used by the EHT are like models that (following Cartwright) are alternative, and possibly contradictory, ways of ``doing the same thing''. If at most one algorithm is correct, then only that algorithm is a reliable epistemic guide. If the correct algorithm---which Weinstein calls the ``God's function''---is one that isn't being used by the EHT, then the results produced by the three EHT imaging algorithms cannot be trusted. In this case the agreement across different imaging algorithms does not increase confidence in the images. 

As we said above, in section \ref{varied_models}, there are ways to respond to this kind of argument in the case of models. In section \ref{auxiliary_conditions}, we described how \textcite{Staley2020} has argued that varying modeling assumptions to represent experimenters’ ignorance is a form of robustness analysis that increases the security of the conclusion. 
Additionally, \textcite{Dethier2020,Dethier2022Unity} explicitly argues for a unified account of robustness across experimental and modeling contexts. This seems to be based on a view of models that treats them as  `epistemic tools' \autocite{Dethier2021models_tools} and centers their reliability or adequacy-for-purpose over their `truth' (see also, e.g., \textcite{Morgan-Morrison1999,Parker2020}. On this view, robustness across varied models can provide confirmation, because model \textit{reports} (like experimental results) are evidence, whose production is apt to be appropriately varied.
In both cases, there is an analogy between models and physical instruments in terms of the role that they play in the context of a particular empirical investigation. It is appropriate to be concerned about the truth or accuracy of the \textit{output} of a model (a model \textit{report}), but two models featuring idealizations may be inaccurate or mutually contradictory while still providing consistent, reliable predictions within some domain of overlapping applicability.   

Even if one does not accept these arguments for the value of robustness across models, another response is available: imaging algorithms are not models---at least not in any sense that would make them vulnerable to Cartwright's criticisms of models. 

The Cartwright-Weinstein critique is based on the idea that at most one model can be true. Every model other than the single ``correct'' model misrepresents the system it is a model of. Here, there is a disanalogy with the EHT imaging algorithms. Unlike models that represent their target in some way, the imaging algorithms used by the EHT Collaboration do not have a representational character; they themselves should not be thought of as reliable representations of the source being imaged. After all, these algorithms reliably reconstruct the appearance of a range of sources, from discs and rings to GRMHD snapshots and ``Frosty the Snowman''. Thus, imaging algorithms cannot plausibly be thought of as true or false, or consistent/inconsistent with one another. Concerns about representational accuracy can only be applied to the images that they produce.

Given this, basing a criticism of the EHT's robustness arguments on \textcite{Cartwright1991}'s critique of robustness is problematic. Instead, there seems to be a strong case for treating imagining algorithms analogously to instruments---as Katie Bouman (personal communication) puts it, they are ``a part of the telescope''. On this view, robustness across data analysis pipelines can be seen as an epistemically valuable way of varying the experiment, even though significant variation in the physical array is not viable. This source of variation in evidence might be seen as a kind of middle ground between variation in instruments and variation in models.     

We will now further situate our discussion of robustness within a broader context of recent philosophical accounts of robustness.

We begin by noting that one of the tacitly accepted benchmarks for a successful philosophical analysis of robustness seems to be the introduction of a new term for some type of robustness argument or analysis. Here we resist that tendency. But it is still useful to see how the details of robustness analysis differ across examples. We will first discuss two astrophysical cases, and then make some connections with recent analyses of robustness in the context of high energy particle physics.

\textcite{Gueguen2020} is skeptical about the value of robustness analyses in astrophysics, which she conceptualizes as equivalent to convergence studies for a class of simulations. She considers a case in which convergent results across a range of cosmological simulations of dark matter haloes failed to increase the security of the results due to the fact that this convergence could be directly attributed to a numerical artifact. In other words, the simulations were not as independent as previously thought, and their agreement was due to shared (and faulty) auxiliary assumptions, which Gueguen interprets as indicating that convergence analysis is insufficient to infer that a particular result is correct.\footnote{Although Gueguen purports to show that robustness is insufficient to establish reliability in the cases she considers, our interpretation of these cases is that they fail to exhibit sufficient robustness; if one has simulations over subregions of the parameter space which converge within that subregion, but are overall inconsistent with simulations from other subregions, then the results are not robust across a sufficient variation in auxiliary assumptions to be considered very reliable. But note that this is demonstrated by their failure to be sufficiently robust---cases where robustness is thus limited cannot be a good guide to the epistemic value of robustness arguments in general.} However, in the EHT case worries about ``failures of independence'' can be to some extent remedied. CLEAN and RML methods do not have much in common (either conceptually or in their implementation), so (in contrast to examples discussed by \textcite{Gueguen2020}) a common numerical artifact responsible for their convergence seems highly implausible.

\textcite{guralp2020evidence} summarizes robustness arguments used in establishing the accelerating expansion of the universe in light of data from supernovae cosmology. Guralp notes that when the SCP team \textcite{perlmutter1999measurements} obtain the best fit for cosmologogical parameters, they rely on one prescription for fitting light-curve, but within that prescription they present 12 different fits, explicitly aiming at ensuring robustness of the result (see table 2 of \textcite{guralp2020evidence} for a summary). Guralp notes (p. 31) that "[t]his is a feasible strategy for the SCP for they have a large number of high-redshift objects to work with". The 12 fits are performed using 4 different analysis techniques, some of which vary the dataset. For example, the ``Effects of Reddest Supernovae'' fitting method uses ``only objects with measured colors, or excludes various objects that are faint''. Another team, High-\textit{z} \textcite{riess1998observational}, uses two light-curve fitting methods (where SCP uses one light-curve fitting method), and also applies ``these methods to different subsets of their objects'' (\textcite{guralp2020evidence} p. 36). The EHT similarly varies the analysis methods (at the imaging stage), but it cannot vary the dataset without further exacerbating the issues resulting from the sparseness of the data. In passing we also note that the blinding stage of the EHT data analysis incorporates what Guralp calls ``robust consistency'' (i.e. that in addition to each teams' individual results being robust, results of two independent teams (SCP and High-\textit{z} converge) inside the single experiment at the blinded analysis stage.

In the context of high energy particle physics, \textcite{karaca2020two} distinguishes between result robustness (RR, defined as ``the invariance of an experimental result across different means of detection'') and procedure robustness (PR, defined as ``the capacity of an experimental procedure to maintain its intended function invariant despite possible variations in its inputs''). Karaca illustrates these using ATLAS as his case study, and overall argues that PR serves as a criterion of validity not only for outcomes of the experiments but also for the experimental procedures themselves. Setting aside the question whether imaging is the appropriate kind of an experimental procedure to which the latter notion applies, we note that the EHT imaging challenges establish exactly this form of reliability for different imaging techniques. The intended function---reliable imaging of a source based on sparse and noise sampling---is well-served by the imaging algorithms for a large set of synthetic datasets.

Finally, \textcite{bogerobustnes} discusses the relationship between trust in experimental results and dependency of those results on computer simulations of systems under investigation, distinguishing between four forms of robustness. In the EHT case general relativistic magnetohydrodynamic simulations play important roles in parameter extraction, because they provide important classes of source models. However, they do not play a central role in the imaging process, which was our main focus. An analysis of the roles simulations play in the EHT would be a useful complementary account to Boge's account of robustness in high energy physics. We also note that the second stage of the EHT imaging process---especially parameter surveys used to establish the relationships between parameter choices for the imaging algorithms and features of the resulting images---have strong parallels to what Boge calls ``parametric robustness'' and ``inverse parametric robustness'' in the context of simulations in high energy physics experiments.

Overall, our analysis of the robustness arguments offered by the EHT makes clear why and how alternative imaging pathways might teach us something new: namely, that the produced image is not an artifact of choices made during a particular analysis pathway.

What we mean by this is that the EHT imaging process produces an image that is demonstrably robust across variation in imaging methods, imaging algorithms, and groups of expert practitioners. It also addresses some key counterfactuals: first, the process demonstrates that \textit{if} the object being observed were different (\textit{e.g.}, a disc or double Gaussian) \textit{then} the output would be correspondingly different. Indeed, at least for the limited variation in ground truth images considered, the methods employed here have been shown to be adequate for reproducing the image. This is additionally strengthened by the observation \autocite[5]{2019EHT_M87_paper2} that 'further tests using EHT observations of quasar calibrator sources (\textit{e.g.}, 3C 279) obtained in the 2017 April observations demonstrate robust structural agreement between RML methods and more traditional CLEAN-based radio imaging techniques.' Second, this imaging process shows a certain stability of outcome when the input remains (more or less) the same. The images for the four observing days are broadly consistent for all three pipelines. The first of these counterfactuals establishes discriminant validation, whereas the second demonstrates convergent validation (in the sense of \textcite{Staley2004robustness}).

In this form of robustness analysis, multiple analysis pathways converge, so \textit{post factum} might be seen as redundant with respect to each other. Indeed, once trust in imaging algorithms has been established (via robustness), the number of independent pipelines used in the future imaging efforts might decrease. However, from an epistemological viewpoint---especially in the context of observing individual sources in astrophysics---they are anything but redundant. This variation in data analysis pipelines provides an important source of variation in the empirical evidence where other sources of variation (e.g., controlling the initial conditions, varying the population, or observing with independent instruments) are not available. Thus establishing the redundancy of these analysis pathways helps demonstrate their reliability and usefulness.

Zooming out to focus on the bigger picture, robustness plays an important role in justifying confidence in the reported conclusions in large experiments (as conceptualized in section \ref{epistemic_situation_large_experiments}), both in astronomy and in other areas. On the surface this might seem paradoxical: robustness is sometimes thought to be valuable only in cases where the experimental conditions are being varied---i.e., the same conclusions are arrived at through a variety of experimental methods. In this paper we have demonstrated how the EHT uses robustness arguments to increase confidence in the images that they produce, despite the practical limitations preventing variation in experimental setup. In doing so, we show that robustness is a crucial component of the epistemology of large scale experiments.

\section*{Acknowledgements}We are grateful to members of the Event Horizon Telescope collaboration, in particular Katie Bouman, Andrew Chael, Shep Doeleman, Peter Galison, and Daniel Palumbo, for numerous conversations about black hole imaging. We also want to thank Erik Curiel, Corey Dethier, Dennis Lehmkuhl, James Nguyen, Frauke Stoll, and audiences at: the BHI Foundations seminar, Bonn's Lichtenberg Group WiP session and the panel on large experiments, and ngEHT HPC Kickoff workshop attendees, for their questions, comments, and feedback. We would also like to thank two anonymous reviewers for this journal for their useful comments and suggestions.

Both authors would like to thank the Volkswagen Foundation for its support in providing the funds to create the Lichtenberg Group for History and Philosophy of Physics at the University of Bonn. This publication is funded in part by the Gordon and Betty Moore Foundation through grant GBMF8273 to Harvard University to support the work of the Black Hole Initiative. This publication was also made possible through the support of a grant from the John Templeton Foundation. The opinions expressed in this work are those of the authors and do not necessarily reflect the views of these Foundations.

\newpage
\printbibliography
\end{document}